\documentclass[12pt]{iopart}

\usepackage{epsfig}

\begin{document}

\title{Hard and soft probe - medium interactions in a 3D hydro+micro approach at RHIC}
\author{S.A.~Bass$^1$, T. Renk$^{2,3}$, J. Ruppert$^4$ and 
	C. Nonaka$^5$}
\address{$^1$ Department of Physics, Duke University, 
             Durham, North Carolina 27708-0305, USA}
\address{$^2$Department of Physics, PO Box 35 FIN-40014 University of 
		Jyv\"{a}skyl\"{a}, Finland}
\address{$^3$Helsinki Institute of Physics, PO Box 64 FIN-00014 University 
		of Helsinki, Finland}
\address{$^4$Department of Physics, McGill University, H3A 2T8, 
		Montreal, Quebec, Canada}
\address{$^5$Department of Physics, Nagoya University, Nagoya 464-8602, Japan}            

\ead{bass@phy.duke.edu}

\submitto{\JPG}

\begin{abstract}
We utilize a 3D hybrid hydro+micro model for a comprehensive and consistent 
description of soft and hard particle production in ultra-relativistic heavy-ion collisions at RHIC. In the soft sector we focus on the dynamics of (multi-)strange 
baryons, where a clear strangeness dependence of their collision rates and freeze-out
is observed. In the hard sector
we study the radiative energy loss of hard partons in a soft medium in the multiple soft scattering approximation. While the nuclear suppression factor $R_{AA}$ does not reflect the high quality of the medium description (except in a reduced systematic uncertainty in extracting the quenching power of the medium), the hydrodynamical model also allows to study different centralities and in particular the angular variation of $R_{AA}$ with respect to the reaction plane, allowing for a controlled variation of the in-medium path-length.
\end{abstract}


Experiments at the 
Relativistic Heavy Ion Collider (RHIC) 
 have established a significant suppression
of high-$p_T$ hadrons produced in central A+A collisions compared to those
produced in peripheral A+A or binary scaled p+p reactions, indicating
a strong nuclear medium effect \cite{Adcox:2001jp,Adler:2002xw}, commonly referred to as {\em jet-quenching}.
Within the framework of perturbative QCD, the leading process of energy 
loss of a fast parton is gluon radiation induced by multiple soft collisions 
of the leading parton or the radiated gluon with color charges in the
quasi-thermal medium \cite{Gyulassy:1993hr,Baier:1996kr,Zakharov:1997uu}.

Over the past two years, a large amount of 
jet-quenching related experimental data has become available 
including but not limited to
the nuclear modification factor $R_{AA}$, the elliptic
flow $v_2$ at high $p_T$ (as a measure of the
azimuthal anisotropy of the jet cross section) and
a whole array of high $p_T$ hadron-hadron correlations.
Computations of such jet modifications have acquired a
certain level of sophistication regarding the
incorporation of the partonic processes involved.
However, most of these calculations have been utilizing
over-simplified models for the underlying soft medium,
e.g. assuming a simple density distribution and its
variation with time. Even in more elaborate setups, most jet
quenching calculations assume merely a one- or two-dimensional Bjorken
expansion.

The availability of a three-dimensional hydrodynamic evolution
code \cite{Nonaka:2006yn} and related hybrid approaches allow
for a much more detailed study of jet interactions in a
longitudinally and transversely expanding medium. 
The variation of the gluon density in these approaches
is very different from  that in a simple Bjorken expansion.
A previous calculation in this direction \cite{Hirano:2002sc,Hirano:2003pw}
estimated the effects of 3-D expansion on the $R_{AA}$.
However, this approach treated the
energy loss of jets in a rather simplified and heuristic manner.
Here, we shall perform a detailed investigation of the
modification  of jets in a  three dimensionally
expanding medium within the BDMPS formalism utilizing quenching weights
as described in \cite{Salgado:2003gb}.
In addition, 3-D hydrodynamic and hybrid models have been
very successful in describing the majority of features
of soft particle production at RHIC (with HBT interferometry being
the sole exception) -- this we shall utilize in order to determine the
medium properties for the jet-quenching calculation.

In this write-up,
we utilize a state-of-the-art fully 
3-D hybrid hydro+micro transport model \cite{Nonaka:2006yn}.
The model employs relativistic 3D-hydrodynamics 
for the early, dense, deconfined stage of the reaction 
and a microscopic non-equilibrium model for the later 
hadronic stage where the equilibrium assumptions are not 
valid anymore. It is capable of self-consistently calculating
the freezeout of the hadronic system, while accounting 
for the collective flow on the hadronization hypersurface 
generated by the QGP expansion. The initial conditions
of the hydrodynamic calculation are tuned to describe the
hadronic data in the soft sector, 
such as hadron yields, spectra, rapidity-distributions as well 
as radial and elliptic flow. Note that while the modeling of the
hadronic stage is of paramount importance for the proper description
of the medium in the soft sector, it does not contribute to the 
jet energy loss.

\begin{figure}[tb]   
\centerline{\epsfig{file=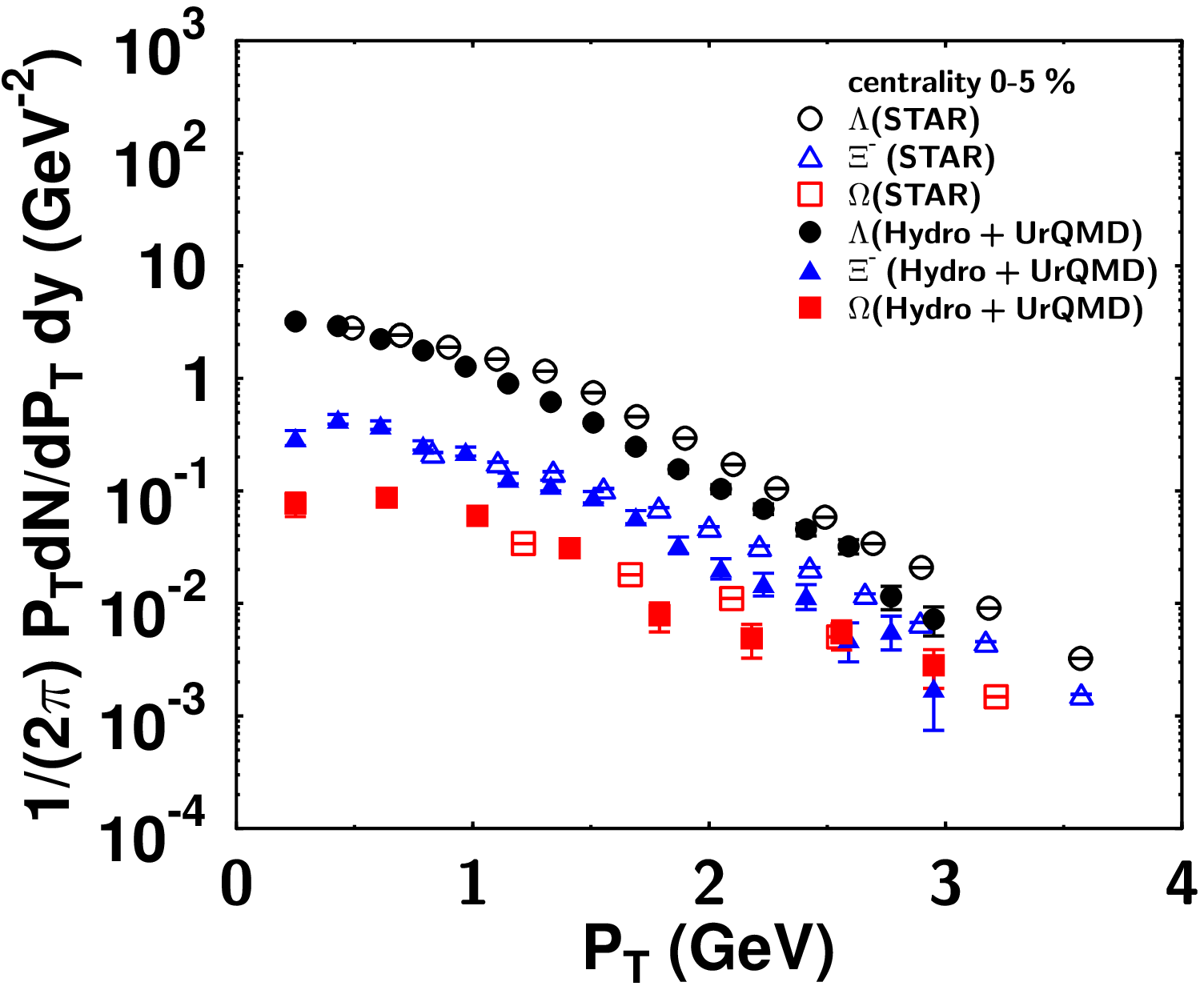,width=8cm}
\epsfig{file=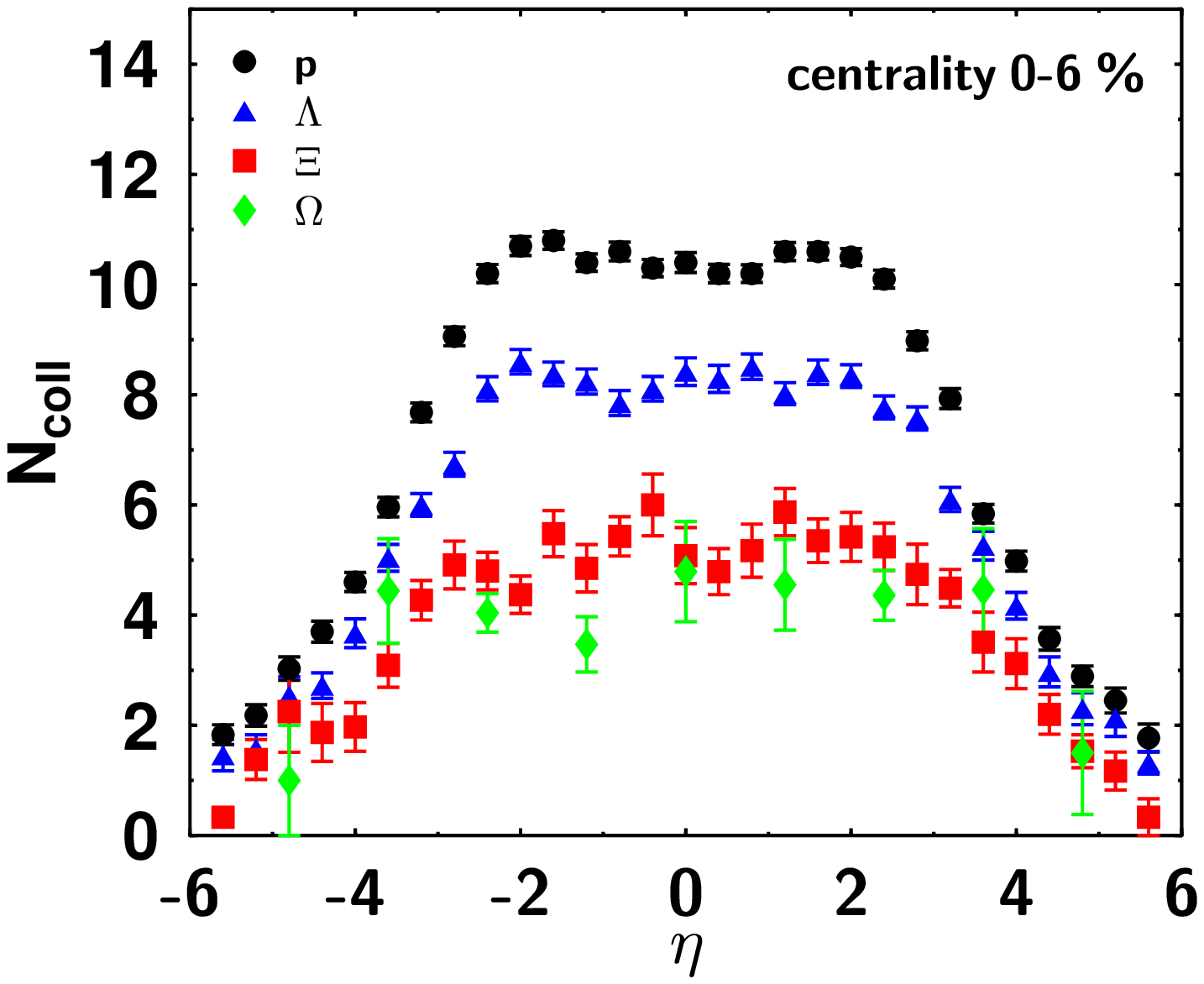,width=8cm}}
\caption{Left: $P_T$ spectra of (multi-)strange baryons calculated
in a hybrid hydro+micro approach compared to STAR data. Right: pseudo-rapidity
distribution of the number of rescatterings for baryons of varying strange quark
content. }
\label{fig1}
\end{figure}

In the left frame of Fig.~\ref{fig1} we analyze the $P_T$ spectra of 
multistrange baryons. Our results show good agreement with   
experimental data for $\Lambda$, $\Xi$, $\Omega$ from the STAR 
collaboration \cite{Estienne:2004di,Adams:2003fy}. 
Recent experimental results suggest that 
at thermal freezeout multistrange baryons exhibit less transverse flow  
and a higher temperature closer to the chemical freezeout 
temperature compared to non- or single-strange baryons
\cite{Estienne:2004di,Adams:2003fy}. This behavior can be understood in terms
of the flavor dependence of the hadronic cross section, which decreases
with increasing strangeness content of the hadron. The reduced
cross section of multi-strange baryons leads to a decoupling from the
hadronic medium at an earlier stage of the reaction, allowing them
to provide information on the properties of the hadronizing QGP less
distorted by hadronic final state interactions 
\cite{vanHecke:1998yu,Dumitru:1999sf,Teaney:2001av}. In microscopic
calculations the early decoupling will manifest itself via a reduced number
of collisions for the respective hadron species.
It should be noted that the analogous behavior has already been observed
in experiments at the CERN-SPS 
\cite{Jones:1996xc,Bormann:1997qa,Appelshauser:1998va,Andersen:1998ph,Andersen:1998vu}.

The pseudo-rapidity dependence of the number of hadronic rescatterings
for different baryon species can be used to corroborate the above findings
and is analyzed in the right frame of  
Fig.~\ref{fig1}, which shows the number of 
collisions of $p$, $\Lambda$, $\Xi$ and $\Omega$ as a function
of $\eta$ at $b=2.4$, 4.5 and 
6.3 fm. The distributions appear to be
similar to that of the particle yield psuedorapidity distribution. 
At midrapidity we find a plateau region extending from $\eta=-3$ to 3, followed
by a steep drop-off to
forward and backward rapidities.
The flavor dependence of the average collision numbers is again clearly seen,
even though we would like to point out that the shapes of the different
distributions are very similar. The large plateau region indicates the 
rapidity domain in which {\em interacting} matter can be found and in which
the application of thermodynamic concepts is viable.

Having determined the properties and dynamics of the 
soft sector, we can now utilize the time-evolution
of the medium provided by our model for the calculation
of jet energy-loss. Thus, our calculation
significantly reduces 
the systematic uncertainties usually associated
with the medium parametrization and allows for a 
precision calculation of all effects associated with hard probe - medium
interactions.
Our calculation follows the BDMPS formalism for radiative energy loss 
\cite{Baier:1996sk} using quenching weights as introduced by
Salgado and Wiedemann \cite{Salgado:2002cd,Salgado:2003gb}.

\begin{figure}[tb]   
\centerline{\epsfig{file=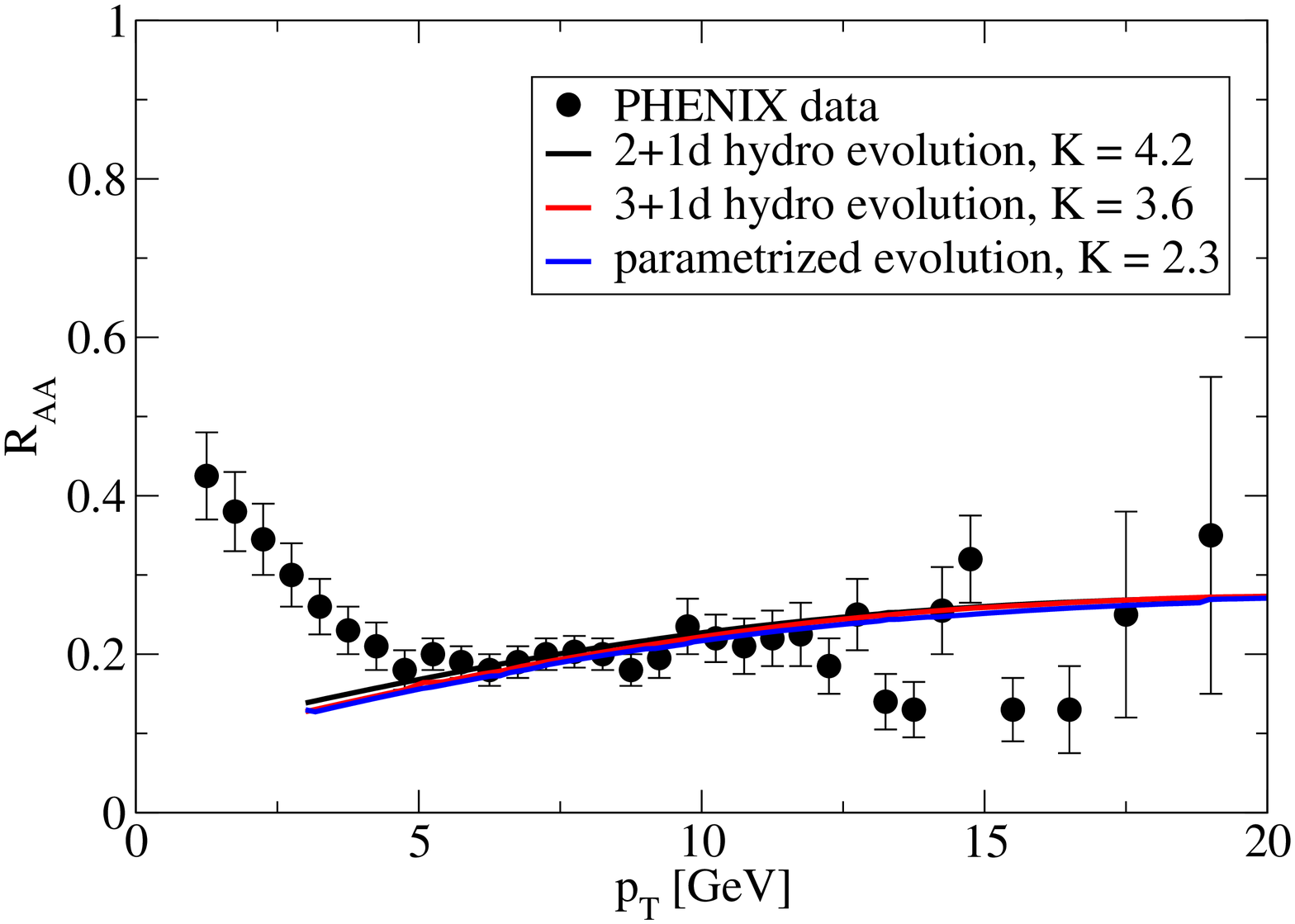,width=8cm}
\epsfig{file=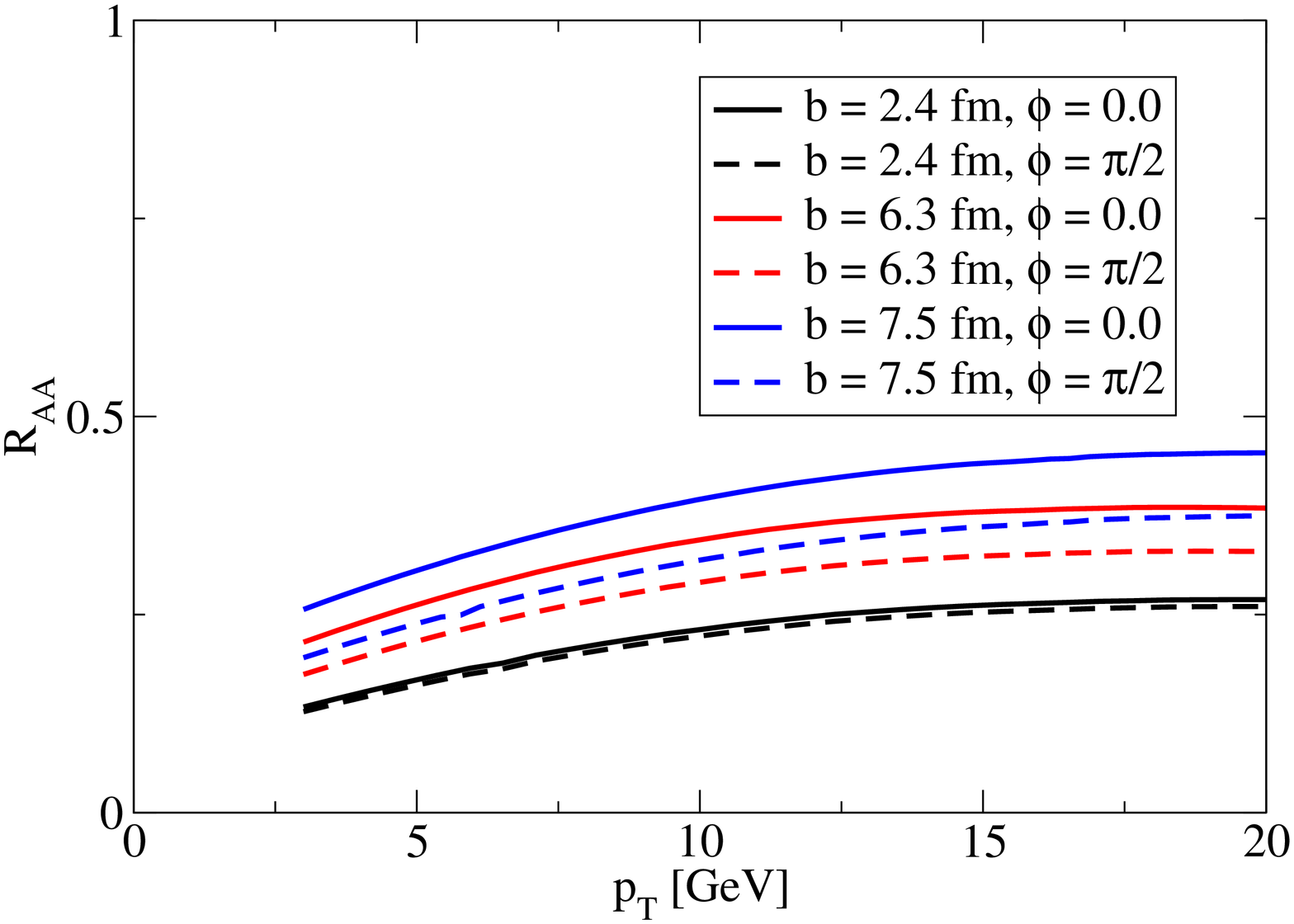,width=8cm}}
\caption{Left: $R_{AA}$ for central collisions as calculated in three different models for the medium evolution with the overall quenching power scale $K$ adjusted to data.  Right: $p_T$ dependence of $R_{AA}$ in plane (solid) and out of plane (dashed) emission at different values of impact parameter ${\bf b}$. }
\label{fig2}
\end{figure}

In \cite{Renk:2006pk,Renk:2006qg} it has been shown that $R_{AA}$ 
for central collisions only constrains a scale, but not the detailed 
functional form of energy loss probability distribution
 $\langle P(\Delta E) \rangle_{T_{AA}}$. In the approach 
outlined above, this is manifest in the parameter $K$ in the 
expression for the local transport
coefficient $\hat{q}(\xi)$:
\begin{equation}
\label{E-qhat}
\hat{q}(\xi) = K \cdot 2 \cdot \epsilon^{3/4}(\xi)
\end{equation}
which then was adjusted 
to the data in central collisions. We illustrate in the left frame of Fig.~\ref{fig2} that 
three different dynamical models, a 2D hydrodynamical evolution \cite{Eskola:2005ue}, 
the 3D hydrodynamical evolution outlined above \cite{Nonaka:2006yn} and a 
parametrized fireball evolution \cite{Renk:2004yv} give almost equal 
descriptions of $R_{AA}$ once the scale parameter is adjusted, albeit they 
require different values of $K$ (the chief reason for this being the 
different longitudinal dynamics). The $\pm 50\%$ spread in the values of {\it K}
for the different models of the medium is a measure for the systematic 
error inherent in the tomographic analysis of jet energy-loss via the
nuclear modification function $R_{AA}$.

However, one may gain predictive power in going 
to collisions at finite impact parameter ${\bf b}$. In particular, the $\phi$ dependence 
of $R_{AA}$ for non-central collisions constitutes a systematic variation 
of path-length within a system with fixed overall scale. The average 
path-length is expected to be smaller for a parton emitted in plane as 
compared to one emitted out of plane, and hence $R_{AA}$ is expected 
to be larger at $\phi=0$ than at $\phi=\pi/2$ with the difference in 
$R_{AA}$ between these angles increasing with the initial asymmetry 
(and hence ${\bf b}$). Using a simple model
for the time-evolution of the medium and collective flow effects, it
has been shown in \cite{Majumder:2006we} that the $\phi$ dependence
of $R_{AA}$ is quite sensitive to the initial gluon density distribution and
temporal evolution of the medium.

Utilizing the previously discussed 3-D RFD model \cite{Nonaka:2006yn}, we 
show the $p_T$  dependence of $R_{AA}$ for emission in plane and out of plane at 
three different impact parameters ${\bf b}$ in the right frame of Fig.~\ref{fig2}.
As expected, $R_{AA}$ grows for more peripheral collisions as there is less soft matter produced to induce energy loss. Moreover, there is a smooth angular variation of $R_{AA}$ observed, reflecting the underlying medium asymmetry. The difference between in-plane and out of plane emission grows with impact parameter, at ${\bf b} = 2.4$ fm there is hardly angular variation whereas at 7.5 fm differences are of order 20\%.

In summary, we  have utilized a 3D hybrid hydro+micro model for a comprehensive and consistent 
description of soft and hard particle production in ultra-relativistic heavy-ion collisions at RHIC. 
In the soft sector we have  focused on the dynamics of (multi-)strange 
baryons, where a clear strangeness dependence of their collision rates and freeze-out
is observed. In the hard sector
we have studied the radiative energy loss of hard partons in a soft medium in the multiple soft scattering approximation. Our analysis should be seen as the starting point for a comprehensive
study of probe-medium interactions treating the hard and soft sector on equal footing.

\ack   
This work was supported in part by an Outstanding Junior Investigator
Award  from the U.~S.~Department of Energy (grant DE-FG02-03ER41239). TR was supported in part
by the Academy of Finland, Project 206024 and JR acknowledges support by the Natural Sciences and Engineering Research Council of Canada.

\section*{References}

\bibliographystyle{iopart-num}

\bibliography{/Users/bass/Publications/SABrefs}

\end{document}